\documentclass[12pt]{iopart}
\usepackage{graphicx}
\usepackage{cite}
\usepackage{url}
\usepackage{subfig}
\RequirePackage{lineno} 
\usepackage{color}
\usepackage{float}
\restylefloat{table}

\begin{document}
\title[]{Simulations of a micro-PET System based on Liquid Xenon}

\author{A. Miceli$^1$, J. Glister$^1$, A. Andreyev $^2$, D. Bryman$^2$, L. Kurchaninov$^1$, P. Lu$^1$, A. Muennich$^1$, F. Retiere$^1$,  and  V. Sossi$^2$}
\address{$^1$ TRIUMF, 4004 Wesbrook Mall, Vancouver V6T 2A3, Canada}
\address{$^2$ Department of Physics and Astronomy, University of British Columbia, 6224 Agricultural Road, Vancouver V6T 1Z1, Canada}

\begin{abstract}
The imaging performance of a high-resolution preclinical microPET system employing liquid xenon as the gamma ray detection medium was simulated. The arrangement comprises a ring of detectors consisting of trapezoidal LXe time projection ionization chambers and two arrays of large area avalanche photodiodes for the measurement of ionization charge  and scintillation light.  A key feature of the LXePET system is the ability to identify individual photon interactions with high energy resolution and high spatial  resolution in 3 dimensions and determine the correct interaction sequence using Compton reconstruction algorithms. The simulated LXePET imaging performance was evaluated by computing the noise equivalent count rate, the sensitivity and point spread function for a point source, and by examining the image quality using a micro-Derenzo phantom according to the NEMA-NU4 standard.  Results of these simulation studies included NECR peaking at 1326 kcps at 188 MBq (705 kcps at 184 MBq) for an energy window of 450 - 600 keV and a coincidence window of 1 ns for mouse (rat) phantoms.  The absolute sensitivity at the center of the field of view was 12.6\%.  Radial, tangential, and axial resolutions of $^{22}$Na point sources reconstructed with a list-mode maximum likelihood expectation maximization algorithm were $\le $ 0.8 mm (FWHM) throughout the field of view. Hot-rod inserts of $<$ 0.8 mm diameter were resolvable in the transaxial image of a micro-Derenzo phantom. The simulations show that a liquid xenon system would provide new capabilities for significantly enhancing PET images.
\end{abstract}
\maketitle

\section{Introduction}

Positron Emission Tomography (PET) is a functional medical imaging technique of increasing importance.  Its power resides in the ability to investigate biological processes that are altered by disease and to trace radio-labeled molecules in organs.  PET imaging can be used for early cancer screening, studying the pathology of illness, and to guide the development of new drugs.   

Recently, several efforts were made to improve the sensitivity and spatial resolution of preclinical PET scanners  by developing scintillation crystal-based detectors capable of measuring depth of interaction \cite{Bergeron09, Tai2003, Sempere2007, Seidel, Yang}. 

We are developing a novel high-resolution preclinical PET system using ionization and scintillation light signals from gamma ray interactions in liquid xenon (LXe). The  Time Projection Chamber (TPC)~\cite{Nygren78} configuration is employed where ionization electrons are collected without gain on electrodes after drifting 11 cm under an applied electric field of 1-3 kV/cm.  Ionization from photon interactions can be localized in 3-D to $<$ 1 mm because electron diffusion is small in LXe. Low diffusion also allows separation of individual photon interactions. Charge collection efficiency is high as long as the level of impurities in the LXe is sufficiently low (ppb level) ~\cite{Chepel1994,Conti2003}.    Photon interactions also produce copious scintillation light in LXe (68000 photons/MeV at zero electric field) with time constants of 2.2 ns and 27 ns, which is detected in our set-up by a set of Large Area Avalanche Photo-diodes (LAAPD) \cite{Moszynski}; scintillation light is used to measure the interaction time with high resolution and contributes to the energy measurement. Furthermore, using both charge and scintillation light, excellent energy resolution ($<4\%$  FWHM at 662 keV) has been reported~\cite{Aprile2007}. LXe can be used to cover large detection volumes with high uniformity over the entire field of view  (FOV) improving the detection sensitivity. Our previous studies on the use of LXe  as a  detection medium in PET were reported in \cite{LXeNIMA}.    The relevant properties of LXe are listed in Table \ref{LXeProperties}.

In this paper, we describe a simulation of a LXe $\mu$-PET scanner and the Compton reconstruction algorithm developed for sequencing multi-interaction events. In addition, we present the simulated imaging performance of the LXePET system including sensitivity, scatter fraction, spatial resolution, and image quality evaluated according to the NEMA standard NU4 ~\cite{NEMA}.

\begin{table} [H]
\caption[]{Properties of liquid xenon.}
\begin{indented}
\item[]
\begin{tabular}{lc}
\hline
Property & Value \\
\hline
\hline
Atomic number                     &           54  \\
Density                                  &            3.1 g/$cm_3$ \\
Boiling point                          &            T = 165 K at 1 atm\\
Melting point                          &            T = 161 K at 1 atm \\ 
Photofraction at 511 keV      &            22 \% \\ 
Attenuation length at 511 keV &           37 mm \\ 
Decay time                         &          2.2ns, 27 ns \\ 
\hline
\end{tabular}
\end{indented}
    \label{LXeProperties}
\end{table}

\section{Simulation Framework}
\subsection{Simulation model}
\begin{figure}[ht]
\centering
\includegraphics[width=4.5in,angle=0]{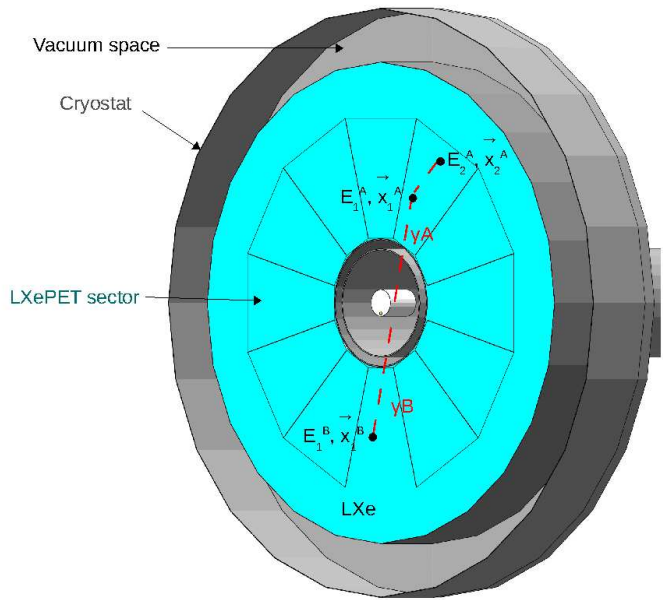}
\caption{Simulated LXe PET system. The cryostat, the twelve LXePET sectors, the inner vessel filled with LXe (blue), and the mouse-like NEMA phantom are illustrated. The red dashed lines indicate a pair of annihilation photons which interact in the LXe. In this figure, photon A interacts twice in the LXePET sector, first via Compton scattering, then via the photoelectric effect. Photon B interacts only once via the photoelectric effect. Energy and 3-D position of each photon interaction are recorded by the TPC. }
\label{fig:PETring}
\end{figure}

\begin{figure}[ht]
\centering
\includegraphics[width=3in,angle=0]{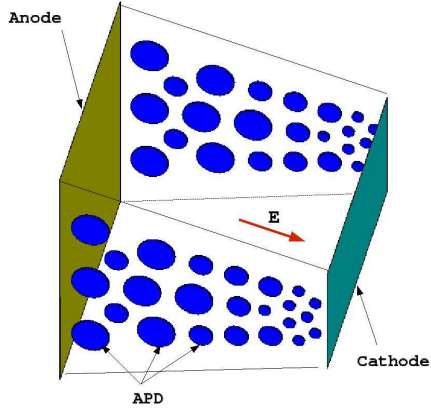}
\caption{APD layout in one of the LXePET sectors. }
\label{fig:apd_configuration}
\end{figure}

Figure ~\ref{fig:PETring}  shows the planned configuration of the LXePET scanner consisting  of twelve trapezoidal sectors arranged in a ring geometry. The inner bore has 10 cm dia. and 10 cm axial length. The liquid xenon is contained in a stainless-steel vessel thermally insulated by a vacuum space. Each sector is a  LXeTPC viewed by two arrays of LAAPDs. The anode and cathode areas are 10 cm x 9.2 cm and 10 cm x 3.2 cm, respectively, and the  drift length is 11.2 cm. Each APD array consists of 7 APDs with 16 mm dia., 9 APDs with 10 mm dia., and 8 APDs with 5 mm dia. Smaller APDs are used in the inner region to enhance the reconstruction where most of the events occur. Figure~\ref{fig:apd_configuration} shows the APD layout in one of the sectors.

The simulation of the LXe prototype was carried out with the Geant4 simulation package \cite{Geant4}. A positron emitter ($^{18}$F or $^{22}$Na depending on the study) was simulated.  Following the decay of the radioisotope, positrons with energy sampled from a continuous distribution of the beta decay process were generated and tracked until annihilation. To  simulate the non-colinearity of the annihilation photons a new process was created and  integrated in Geant4. The new process simulates the positron annihilation in-flight according to the Geant4 annihilation process and replaces the Geant4 annihilation at rest  with a model where the non-zero momentum of the electron-positron pair is taken into account.  The  interactions of the annihilation photons with the phantom and PET scanner were simulated with the low energy package of Geant4.  Energy and 3D position of every photon interaction in the LXe detector were recorded.  The numbers of  ionization charges $N_{e-}^i$ and scintillation photons $S_i$ created  in the interactions were calculated as ~\cite{inpc_conf}:

\begin{equation}
\label{eq:charge_el}
N^i_{e-} =  \frac{(1- Fr^*) \times E_i^{G4}}{15.6eV} 
\end{equation}

 \begin{equation}
\label{eq:light_el}
S^i =  \frac{(\upsilon + Fr^*) \times E_i^{G4}}{15.6eV} 
\end{equation}

where Fr* is the electron-ion recombination fraction, $E_i^{G4}$ is the energy deposited  in the interaction $i$ and $\upsilon$ = 0.2 \cite{Aprile2007} is the ratio of the number of excitons and ion pairs produced.  The electron-ion recombination fraction Fr* varies on an event-by-event basis. It was modeled as a Gaussian function centered at Fr = 0.24 with width $\Delta Fr$ = 0.032 \cite{LXeNIMA}.  
 Electronics and photo-detectors were not simulated directly. Instead, instrumental responses were parameterized in subsequent analyses as described in ~\cite{inpc_conf}.  The parameters used in the simulation are listed in Table  \ref{SimPar}.

\subsection{Event Selection} 
Coincidence events were selected using a two-step procedure. The first stage of the event selection simulated the response of the detector trigger using only the information from the scintillation light. Events producing less than 5000 scintillation photons (corresponding to approximately  180 keV) were rejected.  For each photon of each annihilation pair passing the first selection stage, we calculated the energy from the scintillation light corrected for the solid angle using the information of the position from the charge measurement\cite{inpc_conf} and we used the resulting value to calculate the light-charge combined energy as described in \cite{Aprile2007}.  Events with combined energy  450-600 keV were kept. 
The first interaction points defining the lines of response (LOR) of the selected events were stored in a list-mode format.  The Compton reconstruction algorithm described in section 2.3 was used to find the first interaction point for multi-sites events.

\begin{table}
\caption[]{Simulation parameters.}
\begin{indented}
\item[]
\begin{tabular}{lcc}
\hline
Quantity & Symbol & Value \\ 
\hline
\hline
Recombination factor &  Fr & 0.24  \\
Fluctuation of the e-ion recombination & $\Delta Fr$ & 0.032 \\
Position resolution  & $\sigma_{pos}$ & 0.3 mm \\
Minimum two-hits separation distance & d & 1 mm\\
Electronic noise charge (APD 16mm) & $ENC^{16}_A$ & 5000 e-\\
Electronic noise charge (APD 10mm) & $ENC^{10}_A$ & 2000 e-\\
Electronic noise charge (APD 5mm) & $ENC^{5}_A$ & 500 e-\\
APD quantum efficiency & QE & 80$\%$\\
APD gain & G & 500\\
APD excess noise factor & F & 2.5\\
TPC electronic noise &  $ENC_Q$ &600 e-  \\
Charge detection threshold & $T_C$ &1800 e-\\
\hline
\end{tabular}
\end{indented}
    \label{SimPar}
\end{table}

\subsection{Compton Reconstruction Algorithm}

When a photon interacts in the detector, it can Compton scatter multiple times before being photo-absorbed.  A 511 keV photon is roughly three times more likely to Compton scatter than be photo-absorbed when it first interacts in LXe.  The simplest interaction configuration is the 1-1 case in which the detector registers only 1 discernible interaction point for each of the two photons, corresponding to photo-absorption without scattering.  Practically, however, multi-hit scenarios such as 1-2, 1-3, 2-2, etc. are more common, and must be taken into account, as they contribute to blurring of the image due to ambiguity in the location of the first interaction point.  The goal of the Compton reconstruction algorithm is to sort through all the possible scattering sequences, determine the path that is the most probable, and define the most likely first interaction point and its associated line of response.

For each pair of photons interacting $M-N$ times in the detector, with $M$ representing photon 1 and $N$ photon 2, and $M \leq N$, there are $M!N!$ number of possible interaction sequences.  For each sequence, a LOR check is first performed, determining whether the trajectory passes through the phantom.  Then, if the sequence was found to be viable, Compton kinematics were used to compute a test statistic score associated with the sequence.

The Klein-Nishina formula determines the scattering angle based on the energy deposited:

\begin{equation}
\cos(\theta_{E}) = 1 + mc^2 \times (E^{-1}_{\gamma i} -  E^{-1}_{\gamma i+1})
\end{equation}

where $E_{\gamma i} $ is the photon energy before the $i^{th}$ step given by:

\begin{equation}
E_{\gamma i} = E_{\gamma 1} - \sum_{j=1}^{i-1} dE_j, 
\end{equation} 

m is the electron mass, $\theta_{E}$ is the Compton scattering angle, $dE_{j}$ is the energy deposited at the $j^{th}$ step, and $E_{\gamma 1} = 511  $  keV is the energy of the photon before it  reaches the detector. Alternatively, the scattering angle $\theta_{G}$ based on the position of interaction site is calculated as:

\begin{equation}
\cos(\theta_{G}) = \frac{ \vec{u_{i}} \cdot \vec{u_{i+1}} } { |\vec{u_{i}}||\vec{u_{i+1}}| }
\end{equation}

where $\vec{u_{i}} = (x_{i}-x_{i-1},y_{i}-y_{i-1},z_{i}-z_{i-1}) $.

For each candidate interaction site, we could, in principle, determine if the sequence was the correct one by comparing the scattering angles computed using the energy deposited ($\theta_{E}$) with the observed scattering angles given the geometric distribution of interaction sites ($\theta_{G}$).  In the ideal situation, the difference would be zero.

The ability to resolve the correct sequence, however, depends on the position and energy resolution of the system.  A statistical weighting was used to account for instrumental resolution limits:

\begin{equation}
\chi^{2} = \sum^{N-1}_{i=1}\frac{ \left( \cos(\theta_{E})_{i}-\cos(\theta_{G})_{i} \right)^{2}}{\Delta\cos(\theta_{E})^{2}_{i}+  \Delta\cos(\theta_{G})^{2}_{i}}
\label{eq:5}
\end{equation}

where the error terms are defined as \cite{Aprile2008414}: 

\begin{equation}
\Delta cos(\theta_E)^2_i = m^2c^4 \times \left( \frac{\sigma_{dE_i}^2}{E_{\gamma i}^4} + \sigma_{E \gamma i+1}^2 \times (E^{-2}_{\gamma i} -  E^{-2}_{\gamma i+1})^2\right)
\end{equation}

and 

\begin{equation}
\Delta cos(\theta_G)^2_i = \sigma_{pos}^2 \times (\sigma g_{i,x} + \sigma g_{i,y} + \sigma g_{i,z})
\end{equation}

where 

\begin{equation} 
\vec{\sigma g_i}= \left(  \frac{\vec{u_{i+1}}} {|\vec{u_{i}}| \cdot |\vec{u_{i+1}}|} -   \frac{\vec{u_i} \times cos(\theta_G)} {|\vec{u_{i}}|^2} \right)^2  + \left( \frac{\vec{u_i}} {|\vec{u_{i}}| \cdot |\vec{u_{i+1}}|} -   \frac{\vec{u_{i+1}} \times cos(\theta_G)} {|\vec{u_{i+1}}|^2}  \right)^2
\end{equation}

The error on the energy deposited at the $i^{th}$ step, $\sigma_{dE_i}$, and the error on the photon energy after the interaction step $i$, $\sigma_{E \gamma i+1}$, are given by: 

\begin{equation}
\sigma^2_{dE_i} =ENC_Q^2 + \Delta Fr^2 \times dE_i^2
\end{equation}

\begin{equation}
\sigma^2_{E \gamma i+1} = i \times ENC_Q^2 +  \Delta Fr^2 \times \sum^i_{j=1}dE_j^2
\end{equation}

Finally, the viable sequence with the lowest test statistic score was chosen by the reconstruction algorithm, and the associated LOR defined and recorded.  If no suitable interaction sequence was found, the event was discarded.  This reconstruction technique is similar to the one used in ~\cite{Oberlack2000, Aprile2008414} modified for PET applications. 

\subsection{Pile-up}  \label{sec:pileup}
At high rates,  fast scintillation light signals are used to roughly  (1 $cm^3$) localize the event in order to match correctly the light signal with the slowly drifting charge.  Pile-up of events can occur in the small volume determined by the light localization region and may contribute to the count losses. In order to improve the count rate capability of the LXePET system, a pile-up event recovery method  based on energy balance and proximity  to the light signal was developed. The efficiency of the algorithm was found to be 99 \%, 95\%, and 89\% for 2, 3, and 4-events of simulated pile-up.  The fraction of pile-up events was evaluated by simulating a mouse and a rat phantom filled with water and $^{18}F$.  The time of each decay was simulated using a Poisson distribution. The count rate correction factor for the pileup with the recovery method,  $\epsilon_p$, is given by: 

\begin{equation} \label{eq:pileup}
\epsilon_p  = f_s  + \sum_{k=2}^{4} f_e^k \times  \mu^k
\end{equation} 

where  $f_s$ is the fraction of pile-up free events, $f_e^k$ is the fraction of k-events pileup, and $\mu^k$ is the efficiency of the pile-up recovery method for k-events pileup.

\subsection{Detection Rate Calculation}

The output of the simulation consisted of interaction steps for two types of events: singles where only one of the two photons interacted with the detector and double events where both photons reached the detector. These data are source activity independent and do not contain random coincidence events.  In order to simulate the count rate performance of the LXe detector, the detection rates at different source activities and instrumental parameters, such as dead time and coincidence window, had to be calculated. Also, count losses due to the pile-up of events in the TPC were taken into account. 

The calculation of detection rates was done by Poisson statistical modeling, taking into account the probability of each interaction type, and assuming that only events with exactly two emitted photons detected were selected.  Given the trigger probabilities of detecting zero ($P_0$), one ($P_1$), and two ($P_{2}$) photons from annihilation, the trigger rates for true and scatter events $C_{2,0}$ and for random events $C_{2r,0}$ for a given source activity, A, and coincidence window, $\Delta t$, can be computed:

\begin{equation}
C_{2,0} (A)= \frac{1}{\Delta t}\sum_{k=1}^{\infty} \frac{e^{-\lambda}\lambda^{k}}{k!}P_{2}P_{0}^{k-1},  where ~ \lambda = A * \Delta t ~ and
\label{eq:C20}
\end{equation}

\begin{equation}
C_{2r,0} (A)=   \frac{1}{\Delta t}\sum_{k=2}^{\infty} \frac{e^{-\lambda}\lambda^{k}}{k!}P_{1}^{2}P_{0}^{k-2}
\label{eq:C2r0}
\end{equation}

Coincidence windows of 1, 3, and 6 ns were considered in these studies.  The count rate for true and scatter events $C_{2} (A)$ and for randoms $C_{2r} (A)$ are calculated as: 

\begin{equation}
C_{2} (A)= \frac{C_{2,0}}{1 + C_{total,0}\tau}\epsilon_{2} \times \epsilon_p^2,
\label{eq:C2}
\end{equation}

\begin{equation}
C_{2r} (A)= \frac{C_{2r,0}}{1 + C_{total,0}\tau}\epsilon_{2r} \times \epsilon_p^2
\label{eq:C2r}
\end{equation}

where $\tau$ is the instrumental dead time, $\epsilon_p$ is the count rate correction factor for the pile-up (Eq.\ref{eq:pileup}), and $\epsilon_{2}$ and $\epsilon_{2r}$  are the probabilities of a triggered event to pass the event selection criteria.  $C_{total,0}$ is the total trigger rate including random coincidences.  The ratios $\epsilon_{2}$ and $\epsilon_{2r}$ depend on the combined energy resolution and energy window threshold, as well as on the event reconstruction strategy used.  They are calculated for each data set (simulated true plus scatter data set and random data set) as the number of events which have combined energy within the 450-600 keV energy window and define a LOR which passes through the phantom, divided by the number of triggered events. The random set was generated by combining single unrelated events in pairs.   
The first stage trigger probabilities for zero, one, two photons detection, and 2nd-stage event selection efficiencies are given  in Table~\ref{tab:Int_probs}.  The trigger probability of detecting one or two photons is 60\% for both the mouse and rat phantoms. The probability of detecting two photons depositing more than 180 keV is 22\%, significantly higher for the mouse phantom  than 13\% found for the rat phantom  due to the smaller amount of scattering produced by the mouse phantom.  The amount of scattering is related to the size of the phantom.

Once the two final detection rates were calculated, a rate dependence could be applied to the output of the Geant4 simulation.  This was done by scaling the simulated double (true and scatter) and random events (pair-wise combinations of single events) to obtain  the total detection rate $C_{2}(A) + C_{2r}(A)$.  This scaling approach allowed us to use a single large set of simulation data to compute the behavior of the detector and its performance at various resolution limits and activities without the need to re-simulate under different detector parameters.

\begin{table}
\caption[]{Trigger probabilities for zero, one, two photons ($P_0, P_1, P_2$), and probabilities ($\epsilon_{2}$, $\epsilon_{2r}$) of a triggered event to pass the event selection for non-random and random events.}
\begin{indented}
\item[]
\begin{tabular}{llcc}
\hline
 &  &   Mouse-Like   & Rat-Like \\  
 &  &   Phantom    & Phantom \\  
\hline
\hline
Scenario & $[P_0]$ (\%) & 41.5 & 40.8 \\
& $[P_1]$  (\%) & 36.3 & 46.6 \\
& $[P_{2}]$  (\%) & 22.2 & 12.6 \\
Efficiency & $\epsilon_{2}$ (\%) & 43.3 & 33.5 \\
& $\epsilon_{2r}$ (\%) & 3.39 &  6.26 \\
\hline
\end{tabular}
\end{indented}
    \label{tab:Int_probs}
\end{table}

\subsection{Image reconstruction}   \label{sec:imagerec}
In order to preserve the  high resolution spatial information contained in the data produced by the LXePET scanner, we reconstructed the point source data and the micro-Derenzo phantom with  a list-mode reconstruction algorithm. The main advantages of list-mode data reconstruction over rebinned data reconstruction are preservation of the maximum sampling frequency, and faster reconstruction for low-statistics frames. Data reconstructed with histogram-mode methods are compressed in the axial and radial directions to reduce the sinogram size and to accelerate the reconstruction with a consequent  loss of axial and transaxial resolution \cite{ArmanPhdThesis}. This effect is particularly evident moving away from the axial axis in the transaxial plane. List-mode methods reconstruct the data event-by-event without the need of binning the data into space and time intervals thereby preventing information losses. The information preserving characteristic of list-mode reconstruction methods is particularly useful for high spatial and temporal resolution PET systems ~\cite{Reader}.  The computational time of histogram-mode reconstruction methods depends on the number of line of responses in the sinogram,  whereas reconstruction time of list-mode methods  depends only on number of events recorded. List-mode methods are therefore preferred for high resolution scanners where the number of line of responses can be much higher than the number of recorded events \cite{ArmanPhdThesis}.   List-mode image reconstruction methods are also favorable in time-of-flight PET \cite{Pratx2011}, motion corrected PET \cite{Lamare2007}, and dynamic and gated PET \cite{Rahmim2005}.  We used a 3D list-mode image reconstruction algorithm for PET based on the maximum likelihood expectation maximization (MLEM) approach \cite{Shepp1982}.   As in \cite{Barrett97} and \cite{Parra98} each detected LOR was considered as a unique projection bin with the number of counts in each projection bin $g_i$ equal to 1. Using notations $f_{j}^{n}$ and $f_{j}^{n+1}$ for the intensity vectors in voxel $j$ for step $n$ and the next $n+1$ iteration estimates, the iteration step for the list-mode MLEM algorithm is equal to:

\begin{equation}
f_{j}^{n+1} = \frac{f_{j}^{n}}{{s_{j}}}\sum_{i}{p_{ji}}\frac{1}{\displaystyle\sum \limits_{k}{p_{ik}f_{k}^{n}}}
\end{equation}
where $p_{ij}$ is the value of the system matrix describing the probability that a given emission event $i$ originates from a certain voxel $j$, $s_{j}$ is the sensitivity value for voxel $j$. The list-mode MLEM used \textit{on-the-fly} ray-driven forward and-back projection with bilinear interpolation \cite{Rahmim2004}. We used 20 MLEM iterations for the $^{22}$Na point sources and 100 MLEM iterations for the Derenzo phantom. The voxel size was 0.15 x 0.15 x 0.15 $mm^3$ and the image size was 360 x 360 x 360 voxels. The reconstruction time for point sources (5.5 million LORs on average) was less than 3 hours on an Intel Xeon 2.00 GHz CPU (single core).  The reconstruction speed of the list-mode MLEM algorithm can be further improved by using the ordered subsets (OS) approach \cite{Hudson94} and parallel processing.

\subsection{Simulated Data}
The system performance was evaluated based on the National Electrical Manufacturers Association (NEMA) standards~\cite{NEMA}. The only deviation from the NEMA protocol was the use of a list-mode MLEM reconstruction method instead of FBP reconstruction algorithm for the spatial resolution studies. As explained in section \ref{sec:imagerec}, we used the list-mode MLEM method in order to preserve the  high resolution spatial information of the scanner. 

\textit{Sensitivity}. The sensitivity of the system was determined with a $Na^{22}$ point source embedded in a 1 $cm^3$ acrylic cube.  The source was stepped axially over the axial length of the scanner. The data were rebinned using the SSRB.  \\ 
\textit{Scatter fraction and count rate performance}. The scatter fraction and count rate performance were obtained with a mouse and a rat-like phantoms. The mouse-like phantom was a 25 mm dia. and 70 mm length polyethylene cylinder with a 3.2 mm dia. hole drilled at a radial distance of 10 mm. A simulated 3.2 mm dia. 60 mm long rod was filled with water and $^{18}$F.  
 The rat-like phantom was a 50 mm dia. and 150 mm length polyethylene cylinder with a 3.2 mm dia. hole drilled at a radial distance of 17.5 mm. A simulated 3.2 mm dia. 140 mm long rod was filled with water and the  $^{18}$F.    The data were rebinned using the SSRB.  \\ 
\textit{Spatial resolution}. The spatial resolution was obtained with the $Na^{22}$ point source used for the sensitivity studies. The source was placed at two axial positions 0 and 12.5 mm and five radial positions 0, 5, 10, 15, and 25 mm. The simulated data were reconstructed with the list-mode MLEM iterative method. \\
\textit{Image quality}: The image quality was studied with a micro-Derenzo phantom made from acrylic measuring 40 mm in dia. and 35 mm in length. Arrayed throughout the phantom were cylindrical rods of length 30 mm and diameters 1.6, 1.4, 1.2, 1.0, 0.8 and 0.6 mm.  The rods were  offset radially by 7 mm from the phantom center and filled with water and $^{18}$F.  The rod-to-rod separation was set to twice the rod diameter.  The simulated data were reconstructed with the list-mode MLEM iterative method.

\section{Analysis}
\subsection{Sensitivity}

\begin{figure}[ht]
\centering
\includegraphics[width=3.5in]{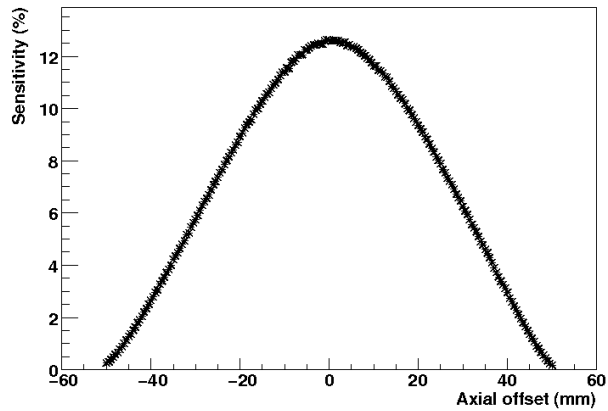}
\caption{NEMA standard sensitivity profile for a $^{22}$Na point source embedded in a 1 cm$^{3}$ acrylic cube, measured at 0.5 mm axial steps.  Energy window: [450,600] keV.}
\label{fig:sensitivity}
\end{figure}

The absolute sensitivity was calculated following the NEMA standard. A simulated $^{22}$Na point source was used for this study.  The point source was stepped axially through the scanner at 0.5 mm steps over an axial length of 100 mm. One million $^{22}$Na decays were simulated at each step.  The total absolute sensitivity for mouse applications was calculated by summing the sensitivity for the sinograms which encompass the central 7 cm. Since the axial extent of the scanner was less than the length of the rat phantom, we calculated the total absolute sensitivity for rat applications summing all the slices, as described in the NEMA standard.  The absolute sensitivity at the center of the field of view (CFOV) for an energy window of [450,600] keV was 12.6 $\%$. The sensitivity profile for all axial steps can be seen in Figure~\ref{fig:sensitivity}.  The  total absolute sensitivity for mouse and rat applications were  9.4 \% and 7.2\%.  The total system sensitivity was 7.2\%. For comparison, typical values of the absolute sensitivity at CFOV range from 3.4\% for the microPET FOCUS-220 with 7.6 cm axial FOV, 250-750 keV energy window, and 10 ns time window  \cite{microPETFocus} to 9.3\% for the Inveon system with 12.7 cm axial FOV, 250-625 keV energy window, and 3.4 ns time window \cite{Inveon} .

\subsection{Scatter Fraction and Count Rate Performance}

The scatter fraction and noise equivalent count rate (NECR) studies were carried out using the rat-like and mouse-like phantoms following the NEMA protocol.  For each phantom 50 million $^{18}$F decays were simulated.  The list-mode simulated true plus scatter data set  was arranged in sinograms (radial bin size 0.3 mm) and oblique slices were combined into 2D projections using the SSRB method with a 1 cm slice thickness. For each sinogram, all pixels located farther than 8 mm from the edges of the phantom were set to zero. The profile of each projection angle was shifted so that the maximum value was aligned with the central pixel of the sinogram. All the angular projections were then summed to generate a sum projection. All counts outside the central 14 mm band were assumed to be scatter counts.  To evaluate the scatter inside the 14 mm central band  we used a linear interpolation. For each slice $i$, the number of scatter counts $C_{scatt,i}$ was given by the total scatter counts in the sinogram (outside and inside the central 14 mm band) divided by the number of pairs in the data set. The total event count  $C_{TOT,i}$ is the sum of the pixels in the projections divided by the number of pairs in the data set.  The scatter fraction is given by

\begin{equation}
SF =\sum_{i=1}^{NSlices} C_{scatt,i}/C_{TOT,i}
\end{equation}

The mouse (rat) scatter fraction was 12.1\%(20.8\%), of which  4.9\% (10.5\%) was due to scatter only and 7.2\% (10.3\%) was due to ambiguities in the Compton reconstruction algorithm. A future paper will deal with reducing the ambiguities. An example of Compton ambiguity involves multi-interaction events where one or both photons interact in only two locations and deposit the same amount of energy.  To calculate the percentage of the scatter fraction due to Compton ambiguities we selected only true events in the simulation data set. 

The random set was arranged in sinograms (radial bin size 0.3 mm) and oblique slices were combined into 2D projections using the SSRB method with a 1 cm slice thickness. The number of random counts $C_{random,i}$ for each slice was the total counts in the random coincidence sinogram within 8 mm from the edges of the phantom divided by the number of pairs in the random set. 

The noise equivalent rate for each slice was calculated as follows, where $C_2(A)$ and $C_{2r}(A)$ are the rates previously calculated:

\begin{center}
\begin{equation}
NECR_i(A) = \sum_{i=1}^{NSlices} \frac{((C_{TOT,i}- C_{scatt,i})\times {C_2(A)})^2}{    C_{TOT,i} \times {C_2(A)}  +   C_{random,i} \times {C_{2r}(A)}   }
\end{equation}
\end{center}

\begin{figure}[ht]
\centering
\includegraphics[width=3.0in]{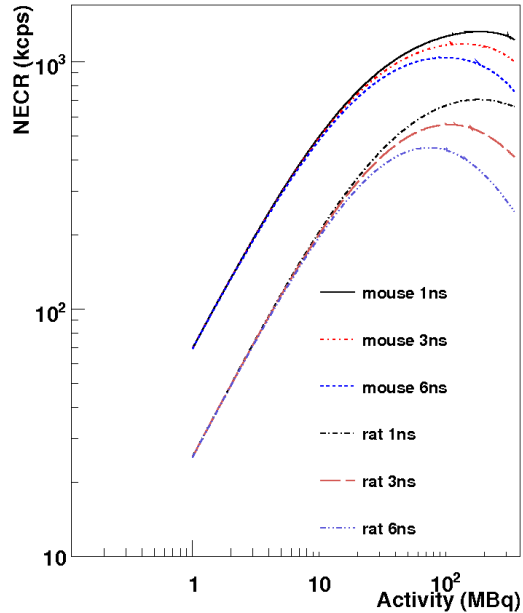}
\caption{NECR vs. total activity, for mouse and rat phantoms and coincidence windows of 1, 3, and 6 ns.  The dead time was 0.2 $\mu$s and the energy window was [450,600] keV.}
\label{fig:NECR_vs_A}
\end{figure}

\begin{figure}[ht]
\centering
\includegraphics[width=3.5in]{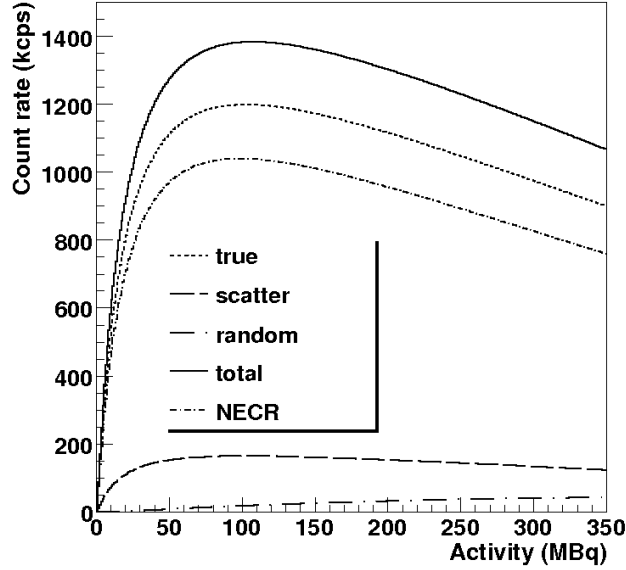}
\caption{True, scatter, random, total counts, and NECR vs. total activity, for mouse-like phantom and coincidence window 6 ns.  The dead time was 0.2 $\mu$s and the energy window was [450,600] keV.}
\label{fig:CountRatePlots}
\end{figure}

The NECR curves for mouse and rat phantoms are plotted in
Figure~\ref{fig:NECR_vs_A} for 1, 3 and 6 ns coincidence windows. The scatter fractions ($SF$), peak true counting rate ($R_{t,peak}$), peak noise equivalent count rate ($R_{NEC,peak}$), activity at which  $R_{t,peak}$  is reached, and activity at which $R_{NEC,peak}$ is reached can be found in Table~\ref{tab:NECR} for mouse and rat phantoms and the three coincidence windows with an energy window of [450,600] keV.   Figure~\ref{fig:CountRatePlots} shows true, scatter, random, total counts, and NECR as a function of activity for the mouse-like phantom with coincidence window 6 ns and dead time 0.2 $\mu$s.  The simulated results show a similar usable range of activity compared with commercial micro-PET systems (1670 kcps at 130 MBq for a mouse phantom, a 350-625 keV energy window and 3.4 ns timing window - Inveon \cite{Inveon}).

\begin{table}
\caption[]{Scatter fraction and count rate performance for rat and mouse phantoms.  The dead time was 0.2 $\mu$s and the energy window was [450,600] keV.}
\begin{indented}
\item[]
\begin{tabular}{ccccccc}
\hline
Phantom & Coincidence & SF (\%) & $R_{t,peak}$ & $R_{NEC,peak}$  &  $A_{t,peak}$  & $A_{NEC,peak}$ \\ 
	  &	Window (ns) &        & (kcounts)           & (kcounts)                   &      (MBq)           &            (MBq)\\
\hline
\hline

Rat     & 1 & 20.8      & 909 &705& 202 & 184\\
        & 3 &      &736 & 558&  122& 108\\
        & 6&      & 605 &450 & 86 &75\\
 
Mouse   & 1       &  12.1          &  1515& 1326& 191 & 188\\
               & 3     &                  & 1359 & 1183&  141& 136\\
               & 6     &                &  1200& 1041&  103& 99\\
\hline
\end{tabular}
\end{indented}
    \label{tab:NECR}
\end{table}

\subsection{Spatial Resolution}

\begin{figure}[t]
\centering
\includegraphics[width=3in, angle=90]{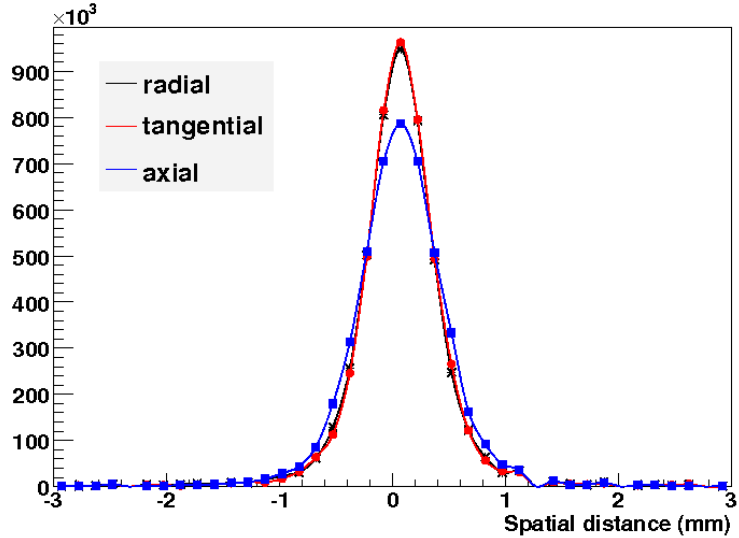}
\caption{Point spread function in radial, tangential, and axial directions of a $^{22}$Na point source at CFOV reconstructed with list-mode MLEM. Radial, tangential, and axial resolutions (FWHM) were 0.6 mm, 0.6 mm, and 0.8 mm.}
\label{fig:PSFatCFOV}
\end{figure}

Spatial resolution was determined using the $^{22}$Na point source with dia. 0.25 mm embedded in a 1 cm$^{3}$ acrylic cube.  A total of 50 million $^{22}$Na decays were simulated and an energy window of [450, 600] keV was used.  It was assumed that the source activity would be low enough that random coincidences could be ignored.  The source was placed at two axial positions: 0 and 12.5 mm. Five radial positions were used for each axial position: 0, 5, 10, 15 and 25 mm.  The data were reconstructed with the list-mode MLEM iterative method (voxel size 0.15 x 0.15 x 0.15 mm, 20 iterations). 
The point spread functions were formed by summing one-dimensional profiles parallel to the direction of measurement and within two FWHM of the orthogonal directions.  The FWHM and FWTM values were calculated through linear interpolation between adjacent pixels at one-half and one-tenth of the peak value in each direction.  The point spread function for a point source at the CFOV is shown in Figure \ref{fig:PSFatCFOV}.

\begin{figure}[t]
\centering
\includegraphics[width=3in, angle=90]{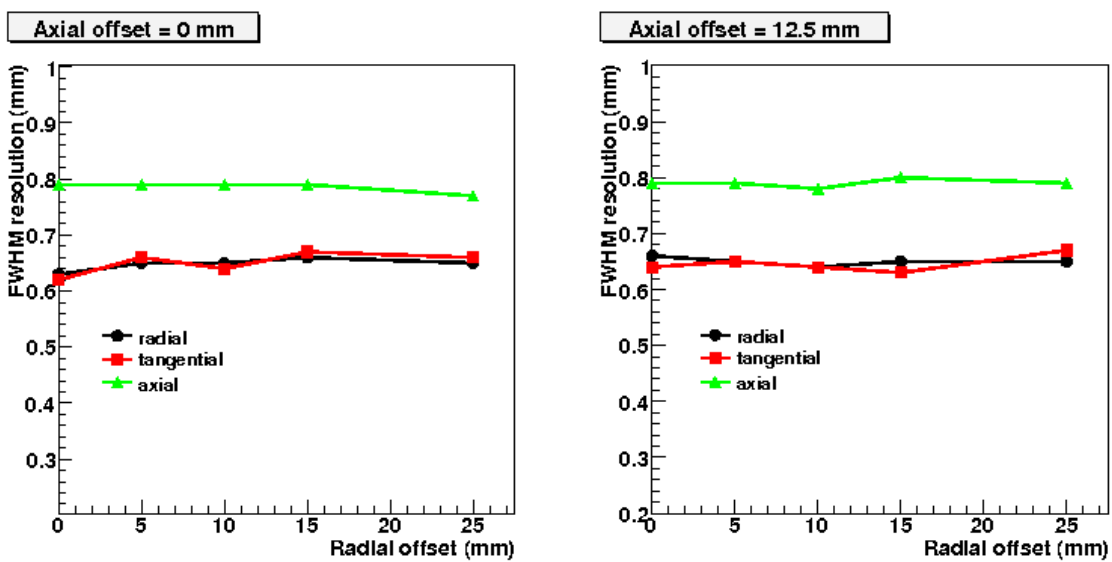}
\caption{Radial, tangential, and axial resolution (FWHM) of a $^{22}$Na point source reconstructed with list-mode MLEM.}
\label{fig:PSF_Resolutions}
\end{figure}

\begin{figure}[h]
\centering
\includegraphics[width=3in, angle=90]{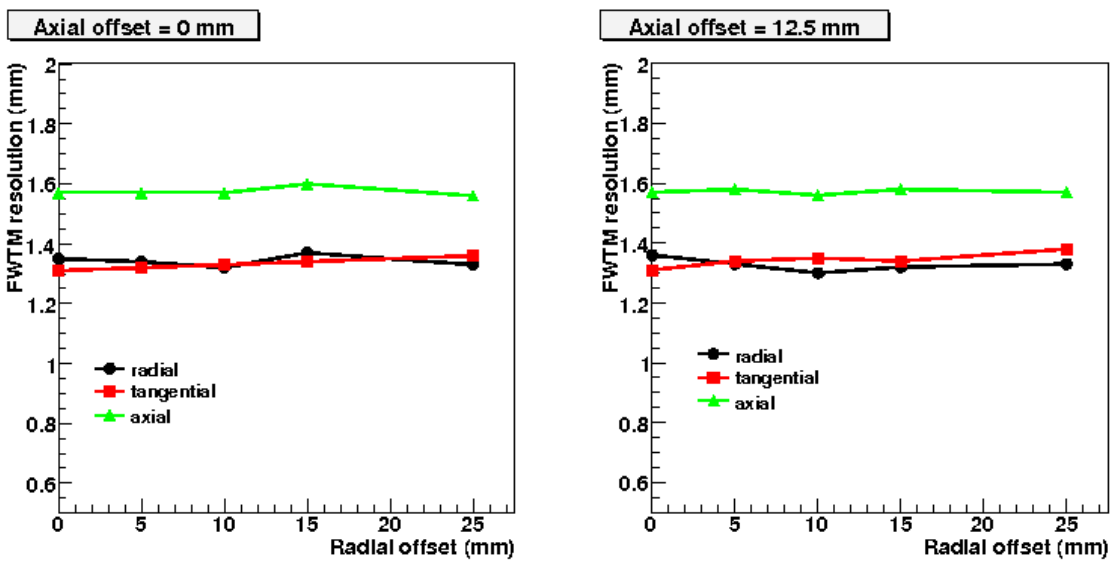}
\caption{Radial, tangential, and axial FWTM of a $^{22}$Na point source reconstructed with list-mode MLEM.}
\label{fig:FWTM}
\end{figure}

Radial, tangential and axial resolutions, reported as FWHM and FWTM, are given  in  Figures \ref{fig:PSF_Resolutions} - \ref{fig:FWTM}.  At the CFOV radial, tangential, and axial  FWHM resolutions of 0.6, 0.6, and 0.8 mm were found.  At 25 mm radial and 12.5 mm axial offset, radial, tangential, and axial  FWHM resolutions were 0.7, 0.7, and 0.8 mm. The results show a uniform resolution $ \leq 0.8$ mm (FWHM) throughout  the FOV in radial, tangential, and axial directions.  At the CFOV, the 2DFBP gave the same results of the MLEM algorithm. For comparison, typical values of spatial resolution for conventional micro-PET systems are 1.3, 1.3, and 1.5 mm (microPET FOCUS-220 \cite{microPETFocus}). Also, the deterioration of the radial resolution towards the periphery of the FOV, which is common for crystals-based preclinical PET systems due to lack of DOI information, is absent in the LXePET. 

\subsection{Image Quality Study}

\begin{figure}[h]
\centering
\includegraphics[width=3.5in]{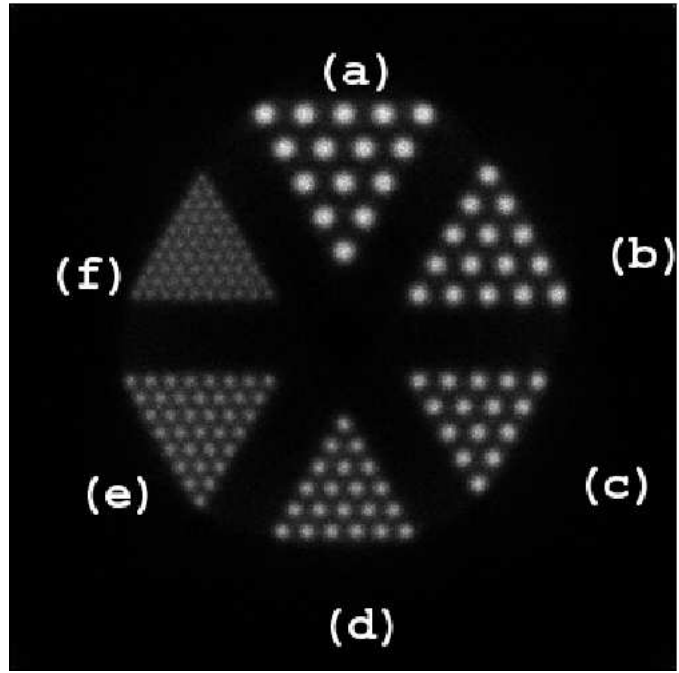}  
\caption{Micro-Derenzo phantom reconstructed using MLEM. Rod-to-rod separation is twice the rod diameter.  Rod diameters: a (1.6mm), b(1.4mm), c(1.2mm), d(1.0mm), e(0.8mm), f(0.6mm).}
\label{fig:Derenzo}
\end{figure}

Figure~\ref{fig:Derenzo} shows a trans-axial slice (thickness 24 mm) of the micro-Derenzo phantom with cylindrical rods of length 30 mm and diameters 1.6, 1.4, 1.2, 1.0, 0.8, and 0.6 mm reconstructed with the list-mode MLEM method (100 iterations).  The voxel size was 0.15 x 0.15 x 0.15 mm. No attenuation or scatter corrections were applied. The source activity was low enough that random coincidences could be ignored. Rods of diameter 0.6 mm  to 1.6 mm are visible. 

\section{Conclusion}

The potential imaging performance of a high resolution liquid xenon
preclinical PET system was evaluated with Monte Carlo simulations. An event
reconstruction algorithm was developed to handle multiple photon scatterings
in liquid xenon, enabling us to refine the lines of response selections and
reduce the event mispositioning introduced by scattered and random events
which result in background noise. Using an energy window [450, 600] keV  which
is possible due to the high energy resolution, the results show that the
LXePET system combines uniform  high resolution radial, tangential, and axial
position measurements throughout the field of view ($\le$0.8 mm FWHM) with high sensitivity (12.6\% at CFOV) and the ability to reject scatter and random coincidences. The scatter fraction was found to be 20.8\%(12.1\%), with associated peak NECR values of 1326 kcps at 188 MBq (705 kcps at 184 MBq) for mouse (rat)-like phantoms. These results show the potentially excellent imaging capabilities of the LXePET systems.  Weighting schemes, where all available data are kept but each LOR is assigned a weight between 0 and 1, and filtering methods based on test statistic score computed with Compton kinematics will be investigated to further decrease noise in the images. Measurements are in progress to demonstrate the performance of the LXePET system described here.

\section*{Acknowledgments}

This work was supported in part by NSERC, CIHR (CHRP Program), the Canada Foundation for Innovation, the University of British Columbia, and TRIUMF which receives founding via a contribution agreement with the National Research Council of Canada.

\section*{References}
\bibliographystyle{unsrt}
\bibliography{LXe_PhysMed}

\end{document}